# Basic Network Creation Games with Communication Interests[*] [†]

Andreas Cord-Landwehr
Heinz Nixdorf Institute &
Computer Science Department
University of Paderborn (Germany)
andreas.cord-landwehr@upb.de

Martina Hüllmann
Computer Science Department
University of Paderborn (Germany)
martina.huellmann@upb.de

Peter Kling
Heinz Nixdorf Institute &
Computer Science Department
University of Paderborn (Germany)
peter.kling@upb.de

Alexander Setzer
Heinz Nixdorf Institute &
Computer Science Department
University of Paderborn (Germany)
asetzer@upb.de

**Abstract**

Network creation games model the creation and usage costs of networks formed by a set of selfish peers. Each peer has the ability to change the network in a limited way, e.g., by creating or deleting incident links. In doing so, a peer can reduce its individual communication cost. Typically, these costs are modeled by the maximum or average distance in the network. We introduce a generalized version of the *basic network creation game* (BNCG). In the BNCG (by Alon et al., SPAA 2010), each peer may replace one of its incident links by a link to an arbitrary peer. This is done in a selfish way in order to minimize either the maximum or average distance to all other peers. That is, each peer works towards a network structure that allows himself to communicate efficiently with *all* other peers. However, participants of large networks are seldom interested in all peers. Rather, they want to communicate efficiently only with a small subset of peers. Our model incorporates these (communication) *interests* explicitly. In the MAX-version, each node tries to minimize its maximum distance to nodes it is interested in. Likewise, the goal of each node in the AVG-version is to minimize the corresponding average distance.

Given peers with interests and a communication network forming a tree, we prove several results on the structure and quality of equilibria in our model. For the MAX-version, we give an upper worst case bound of $\mathcal{O}\left(\sqrt{n}\right)$ for the private costs in an equilibrium of $n$ peers. Moreover, we give an equilibrium for a circular interest graph where a node has private cost $\Omega\left(\sqrt{n}\right)$, showing that our bound is tight. This example can be extended such that we get a tight bound of $\Theta\left(\sqrt{n}\right)$ for the price of anarchy. For the case of general communication networks we show the price of anarchy to be $\Theta\left(n\right)$. Additionally, we prove an interesting connection between a maximum independent set in the interest graph and the private costs of the peers. For the AVG-version, we give a linear lower bound on the worst case private costs in an equilibrium.

[*]This work was partially supported by the German Research Foundation (DFG) within the Collaborative Research Centre "On-The-Fly Computing" (SFB 901) and by the Graduate School on Applied Network Science (GSANS).

[†]An extended abstract of this paper has been accepted for publication in the proceedings of the 5th International Symposium on Algorithmic Game Theory (SAGT), available at www.springerlink.com.

# 1 Introduction

In a network creation game (NCG), several selfish players create a network by egoistic modifications of its edges. One of the most famous NCG models is due to Fabrikant et al. [6]. Their model intends to capture the dynamics in large communication and computer networks built by the individual participants (peers, players) in a selfish way: participants try to ensure a network structure supporting their own communication needs whilst limiting their individual investment into the network. Since the players do not (necessarily) cooperate, the resulting network structure may be suboptimal from a global point of view. The analysis of the resulting structure and its comparison to a (socially) optimal structure is a central aspect in the analysis of network creation games.

In the original model by Fabrikant et al., players may buy (or create) a single edge for a certain (fixed) cost of $\alpha > 0$. Their goal when buying edges is to improve the network structure with respect to their individual communication needs. There are typically two ways to formalize the corresponding communication cost of a single peer: the maximum distance to any other node in the network (local diameter) or the average distance to all other nodes in the network. We refer to the different variants by MAX-version and AVG-version, respectively. Alon et al. [2] introduce a slightly simpler model, called *basic network creation games* (BNCG). Here, they drop the cost parameter $\alpha$. Instead, they limit the possible ways in which peers may change the network by restricting them to edge swaps: a peer may only replace one of its incident edges with a new edge to an arbitrary node in the network. Since peers are assumed to be selfish, only edge swaps (also including simultaneous swapping of several edges at once) that improve the private communication cost of the corresponding peer are considered. In a *swap equilibrium*, no player can decrease its communication cost by an edge swap. This simpler variant of network creation games has the advantage of an equilibrium notion that allows for a polynomially computable best response of the players. Moreover, it still captures the inherent dynamic character and difficulty of communication networks formed by selfish participants, while avoiding the quite intricate dependence on the parameter $\alpha$ (see related work).

Our work generalizes the BNCG model of Alon et al. by introducing the concept of *interests*. In real communication networks, like the Internet, participants are typically only interested in a small subset of peers rather than the complete network. Thus, each participant has a certain set of peers it is interested in. Now, instead of trying to minimize the maximal or average distance to *all* other nodes, the individual players consider only the corresponding distances to nodes they are interested in. The main part of our analysis focuses on tree networks (which are preserved by the peers). Especially, we show that tree networks perform much better than general networks with respect to the price of anarchy. To avoid networks to become disconnected (note that in a BNCG peers want to communicate with all other peers and hence never disconnect the network), we restrict the peers to swaps that preserve connectivity of the network. This restriction is valid from a practical point of view, where a lost network connectivity is to be avoided, since re-connecting a network causes high or even unpredictable costs (if it is possible at all). Moreover, if you consider that interests of the peers may change over time, it is also important for each single selfish peer to sustain connectivity.

## 1.1 Model & Notions

An instance of the *basic network creation game with interests (I-BNCG)* is given by a set of $n$ players (peers, nodes) $V = \{v_1, v_2, \ldots, v_n\}$, an initial *connection graph* $G = (V, E)$, and an *interest graph* $G_I = (V, I)$. We use $I(v) := \{u \in V \mid \{v, u\} \in I\}$ to refer to the neighborhood of a player $v$ in the interest graph and denote them as the *interests of* $v$. Both the connection graph and the interest graph are undirected. Thus, interests are always mutual. The connection graph represents the current communication network and can change during the course of the game. We consider only instances where the (initial) connection graph is a tree, whereas the interest graph $G_I$ may be an arbitrary and not necessarily connected graph. Each player is assumed to have at least one interest. Given a current connection graph, we study two different ways to formalize the private communication costs of nodes: the *MAX-version* and the *AVG-version*. In the first, the private cost $c(v) := \max\{d(v,u) \mid u \in I(v)\}$ of a node $v \in V$ is defined as the maximum distance from $v$ to the nodes it is interested in (its *interests*). In the second, we define $c(v) := \sum_{u \in I(v)} \frac{d(v,u)}{|I(v)|}$ as the average distance to its interests. Here, $d(v,u)$ denotes the (shortest path) distance between $u$ and $v$ in the connection graph.

To improve its private cost, a player $u$ may perform *edge swaps* in the connection graph: replace an incident edge $\{u,v\}$ with a new edge $\{u,w\}$ to an arbitrary player $w \in V$, written as $u : [v \to w]$. We refer to a single as well as to a series of simultaneously executed edge swaps of a player $u$ as an *improving step*, if $u$'s private cost decreases. A player is only allowed to perform an improving step if the connection graph stays connected. If no player can perform an improving step, we say the connection graph is in a *MAX-equilibrium* or *AVG-equilibrium*, respectively. See Figure 1 for an example.

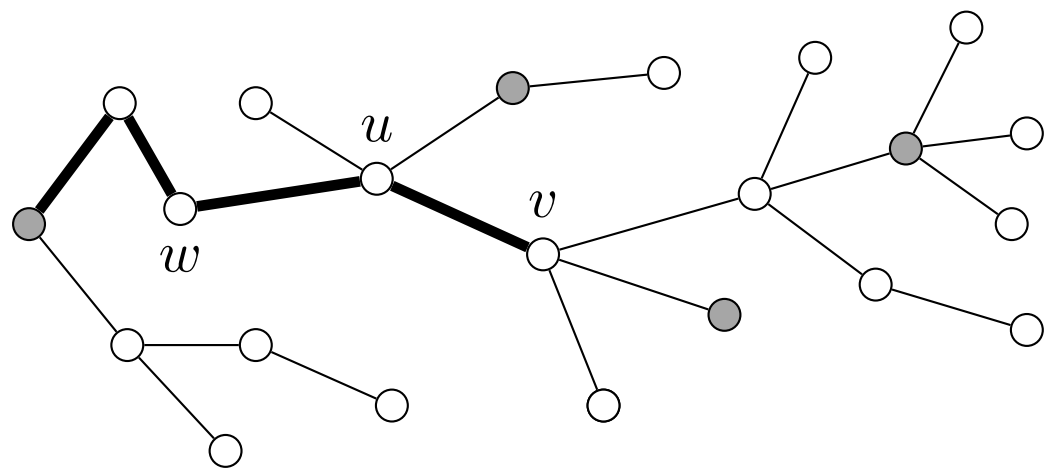

(a) Connection graph where node $v$ is not in a MAX-equilibrium, $c(v) = 4$.

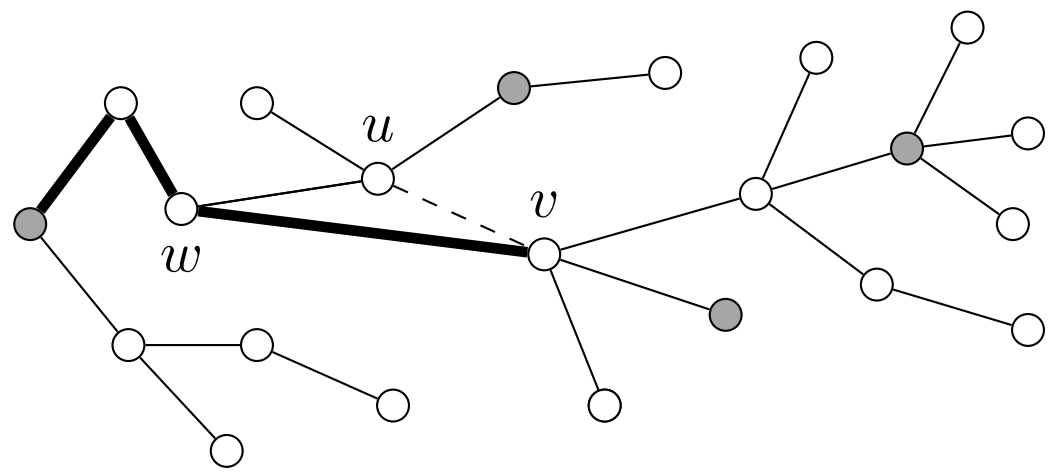

(b) After swap $v : [u \to w]$ the private cost of $v$ is $c(v) = 3$.

Figure 1: MAX-version example of a connection graph. The gray nodes denote $I(v)$, the thick line a longest shortest path from $v$ to a node in $I(v)$. In (a) we have private cost $c(v) = 4$. By the swap in (b) the private cost decreases to $c(v) = 3$.

The overall quality of a connection graph $G$ is measured by the *social cost* $c(G) := \sum_{v \in V} c(v)$ as the sum over all private costs. Our goal is to analyze the structure and social cost of worst case swap equilibria and compare them with a general optimal solution (which is not necessarily a swap equilibrium). As usual in algorithmic game theory, we use the ratio of these two values (*price of anarchy*, see Section 2.3) for this comparison [8]. Note that if the interest graph equals the complete graph, our I-BNCG coincides with the BNCG by Alon et al. [2].

## 1.2 Related Work

Network creation games combine two crucial aspects of modern communication networks: network design and routing. Many such networks consist of autonomous peers and have a highly dynamic character. Thus, it seems natural to use a game theoretic approach to study their evolution and behavior. Given the possibility to change the network structure (buy bandwidth, create new links, etc.), peers typically try to improve their individual communication experience. The question whether this selfish behavior results in an overall good network structure constitutes the central question of the study of network creation games as introduced by Fabrikant et al. [6]. In their model, the authors use a fixed cost parameter $\alpha > 0$ representing the cost of buying a single edge. The players (nodes) in such a network creation game can buy edges to decrease their local communication cost (the average distance to all other nodes in the network). Each player's objective is to minimize the sum of its individual communication cost and the money spent on buying edges. In their seminal work, the authors proved (among other things) an upper bound of $\mathcal{O}(\sqrt{\alpha})$ on the price of anarchy (PoA) in the case of $\alpha < n^2$. Albers et al. [1] proved a constant PoA for $\alpha = \mathcal{O}(\sqrt{n})$ and the first sublinear worst case bound of $\mathcal{O}(n^{1/3})$ for general $\alpha$. Demaine et al. [5] were the first to prove an $\mathcal{O}(n^{\varepsilon})$ bound for $\alpha$ in the range of $\Omega(n)$ and $\mathrm{o}(n \lg n)$. Furthermore, Demaine et al. introduced a new cost measure for the private cost, causing the individual nodes to consider their maximum distance to all remaining nodes instead of the average distance. For this variant they showed that the PoA is at most 2 for $\alpha \geq n$, $\mathcal{O}\left(\min\{4^{\sqrt{\lg n}}, (n/\alpha)^{1/3}\}\right)$ for $2\sqrt{\lg n} \leq \alpha \leq n$, and $\mathcal{O}\left(n^{2/\alpha}\right)$ for $\alpha < 2\sqrt{\lg n}$. Recently, Mihalák and Schlegel [10] could prove that for $\alpha > 273 \cdot n$ all equilibria in the AVG-version are trees (and thus the PoA is constant). If $\alpha > 129$, the same result applies to the MAX-version.

While network creation games as defined by Fabrikant et al. and their variants seem to capture the dynamics and evolution caused by the selfish behavior of peers in an accurate way, there is a major drawback of these models: most of them compute the private communication cost of the peers over the *complete* network. Given the immense size of such communication networks, this seems rather unrealistic. Typically, participants want to communicate only in small groups, with a small subset of other participants they know. To the best of our knowledge, the only other work taking this into account is due to Halevi and Mansour [7]. They introduce a concept similar to our interests (see model description). For the objective of minimizing the average distance of a peer to its interests, Halevi and Mansour proved the existence of pure nash equilibria for almost all $\alpha$ (in particular, $\alpha \leq 1$ and $\alpha \geq 2$). For general $\alpha$ they provided an upper bound of $\mathcal{O}(\sqrt{n})$ for the PoA. In the case of $\alpha$ or $d$ constant (where $d$ denotes the average degree in the interest graph) or $\alpha = \mathcal{O}(nd)$, Halevi and Mansour upper bounded the PoA by a constant. Furthermore, the authors provided a family of problem instances for which the PoA is lower bounded by $\Omega(\log n / \log\log n)$.

As can be seen from the results stated above, the results in all variants of these network creation games largely depend on the cost parameter $\alpha$. Moreover, as has been stated in [6], computing a player's best response for these models is NP-hard. This observation leads to a new, simplified formalization by Alon et al. [2], trying to capture the crux of the problem without the burden of this additional parameter. They introduce basic network creation games (BNCG), where players no longer have to pay for edges. Instead, their possible actions are limited to so called *improving edge swaps*: exchanging a single, incident edge with an edge to some arbitrary node in the network which improves the node's private cost. Other than that, the general problem stays untouched, especially the private cost function (average distance or maximum distance to all other nodes). Best responses in the resulting game turn out to be polynomially computable. Restricting the initial network (connection graph) to trees, they show that the only equilibrium in the AVG-version is a star graph. Without restrictions, all swap equilibria are proven to have a diameter of $2^{\mathcal{O}\left(\sqrt{\lg n}\right)}$. For the MAX-version, the authors prove a maximum diameter of 3 if the resulting equilibrium is a tree. Furthermore, the authors show the existence of an equilibrium of diameter $\Theta(\sqrt{n})$. Our model is a direct generalization of these BNCGs, introducing the concept of interests that model the fact that participants are seldom interested in the complete network.

Up to now, the only other work on BNCGs we are aware of is due to Lenzner [9]. He studies the dynamics of the AVG-version of BNCGs and proves for the case of tree connection graphs a convergence to pure equilibria. Moreover, he proves that any sequence of improving edge swaps converges in at most $\mathcal{O}\left(n^3\right)$ steps to a star equilibrium.

### 1.3 Our Contribution

In our work, we introduce a generalized class of the BNCG by taking the different interests of individual peers into account. We analyze the structure and quality for the case that the initial connection graph is a tree. For the MAX-version, we derive a worst case upper bound of $\mathcal{O}(\sqrt{n})$ for the private costs of the individual players in an equilibrium. Thereto, we introduce and apply a novel combinatorial technique that captures the structural properties of our equilibria (see MAX-arrangement, Definition 2.4). Further, for interest graphs with a maximum independent set of size $M \leq \sqrt{n}$ (e.g., the clique graph with $M = 1$), we can improve the private cost upper bound to $\mathcal{O}(M)$. Using a circular interest graph, we construct an equilibrium with a player having private cost $\Omega\left(\sqrt{n}\right)$, showing that our bound is tight. By extending this construction, we are able to prove a tight bound of $\Theta\left(\sqrt{n}\right)$ on the price of anarchy (ratio between the social cost of a worst case equilibrium and an optimum [8]). Using a star-like connection graph, we show the existence of a MAX-equilibrium with small social cost, yielding a *price of stability* (ratio between the social cost of a best case equilibrium and an optimum [4, 3]) of at most two for an I-BNCG. For the case of an I-BNCG featuring a general connection graph (instead of a tree), we show that the price of anarchy is $\Theta\left(n\right)$. Additionally, for the AVG-version we show a private cost lower bound for individual players of $\Omega\left(n\right)$.

# 2 Maximum Distance Games with Interests

In this section we consider the MAX-version. For this, we show a tight worst case upper bound for the private cost of every node in a MAX-equilibrium of $\Theta\left(\sqrt{n}\right)$ (with $n$ being the number of nodes) as well as the same upper bound for the price of anarchy. The price of stability we can limit to be at most two. For general connection graphs (that may contain cycles) we provide an instance with social cost $\Omega\left(n^2\right)$, yielding a price of anarchy of $\Theta\left(n\right)$.

## 2.1 Private Cost Upper Bound

In the following we prove the private cost upper bound as stated below:

**Theorem 2.1** *Let $I$ be a set of interests and $G = (V, E)$ a corresponding tree in a MAX-equilibrium, $n := |V|$. Then, for all $v \in V$ we have $c(v) \in \mathcal{O}\left(\sqrt{n}\right)$.*

*Outline of the proof:* We consider a tree network in a MAX-equilibrium and take an arbitrary node with maximal private cost among all nodes. Starting with this node, we define a special node sequence, called MAX-arrangement, that will contribute the following properties: each two successive nodes of the sequence are interested in each other and every node of the sequence is "far away" from all previous nodes of the sequence. We will prove that such a sequence necessarily exists and that its length is proportional to the private cost of the starting node.

In detail, we prove with Lemma 2.6 and Lemma 2.7 that a shortest path traversal of a MAX-arrangement in the connection graph uses each edge of the tree at most twice and by this limits its length. Lemma 2.8 constructively shows that given a node with maximal private cost, there always exists a MAX-arrangement starting with this node and ending with a node with a private cost of 3. Lemma 2.5 gives us that the number of nodes in this MAX-arrangement is proportional to the maximal private cost of the first node. Comparing the maximum private cost of a node with the length of a shortest path traversal of any corresponding MAX-arrangement gives us the upper bound.

**Remark 2.2** *Note that in a MAX-equilibrium, each node $v$ with $|\,I(v)| = 1$ has $c(v) = 1$. Hence, for a node $v'$ with $c(v') > 1$ it holds $|\,I(v')| > 1$.*

**Lemma 2.3 (T-configuration)** *Let $I$ be a set of interests and $G = (V, E)$ a corresponding tree in a MAX-equilibrium, $v \in V$ with $|\,I(v)| \geq 2$. Then there exist nodes $x, y \in I(v)$ such that $|d(x, v) - d(v, y)| \leq 1$ and $v$ is connected by at most one edge to the shortest path from $x$ to $y$ and $c(v) = d(v, x)$.*

**Proof.** Let $v \in V$ with $|\,I(v)| \geq 2$ and let $x \in I(v)$ with $d(v, x) = c(v)$. Assume for contradiction that all nodes $x' \in I(v) \setminus \{x\}$ are at distance $d(x', v) \leq c(v) - 2$ from $v$. Consider the shortest path $v \to v_1 \to v_2 \to \ldots \to x$ to node $x$. In this case $v$ can reduce its private cost by the swap $v : [v_1 \to v_2]$ since this swap improves $v$'s distance to $x$ by 1 but increases the distances to every node in $I(v) \setminus \{x\}$ by at most 1. Hence, this contradicts the graph being in a MAX-equilibrium.

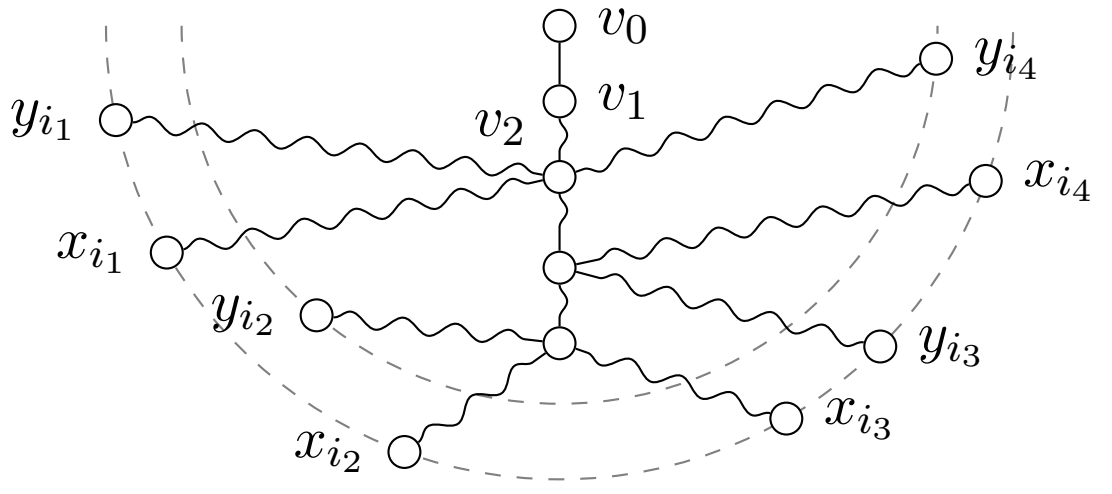

Figure 2: Visualization for Lemma 2.3. Node $v_0$ can perform improving swap $v_0 : [v_1 \to v_2]$.

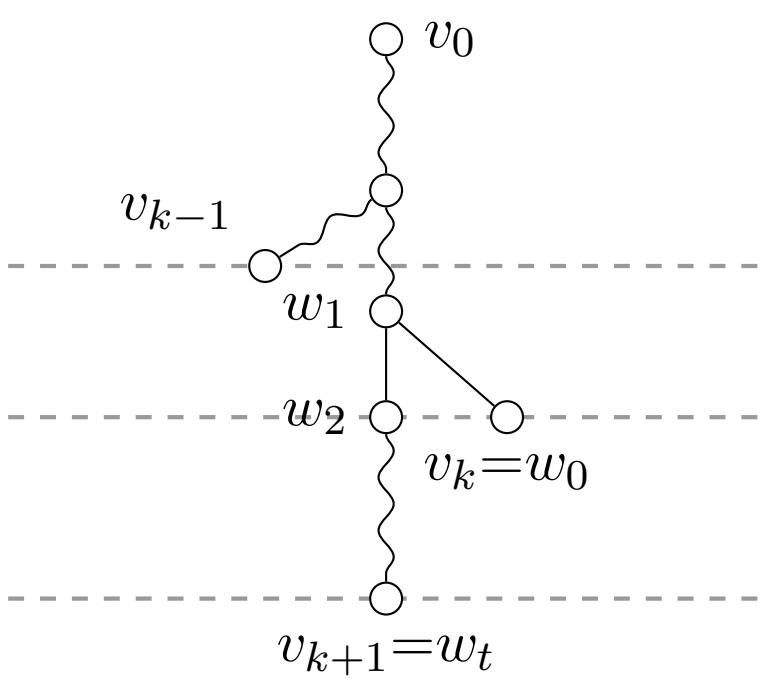

Figure 3: Visualization for the proof of Lemma 2.7. Edge $\{w_0, w_1\}$ is used two times.

We now consider all pairs $(x_i, y_i) \in I(v) \times I(v)$ for that hold $d(v, x_i) = c(v)$ and $d(v, y_i) \geq c(v) - 1$. Let us assume that $v$ is connected to each shortest path from $x_i$ to $y_i$ by at least two edges that do not lie on that path. (See Figure 2 for a visualization.) Thus, $v$ is not located on the shortest path from $x_i$ to $y_i$. This implies that in the graph $G \setminus \{v\}$ for each pair $(x_i, y_i)$ there exists a connected component containing both nodes $x_i, y_i$. Since each two nodes at distance exactly $c(v)$ form such a pair, all nodes of $I(v)$ at distance exactly $c(v)$ must be located in the same connected component, which then gives for every pair $(x_i, y_i)$ that both nodes are contained in the same connected component. Hence, all nodes $x' \in I(v)$ at distance $d(x', v) \geq c(v) - 1$ from $v$ are in the same connected component and by the two edges distance constraint, there must be a path $v \to v_1 \to v_2$ that is a subpath of every path from $v$ to every node $x_i$ and $y_i$. Hence, $v$ can perform the improving swap $v : [v_1 \to v_2]$ (cf. Figure 2). This swap decreases the distance to all nodes $x_i, y_i$ by one and increases each distance to other nodes (i.e., nodes $w \in I(v)$ with $d(w, v) \leq c(v) - 2$) by at most one and hence contradicts $G$ being in a MAX-equilibrium. □

**Definition 2.4 (MAX-arrangement)** *Let $v_0 \in V$ and $v_1 \in I(v_0)$ such that $d(v_0, v_1) = c(v_0)$. Consider a sequence of nodes $v_0, \ldots, v_m$ with $v_i \in I(v_{i-1})$, $i = 1, \ldots, m$, with private costs $c(v_i) > 3$ for $i = 0, \ldots, m-1$ and $c(v_m) = 3$. We call this sequence a* MAX-arrangement *if for all $i = 2, \ldots, m$ it holds:*

$$v_i = \operatorname*{argmax}_{v_i \in I(v_{i-1})} \left\{ d(v_{i-2}, v_i) \;\middle|\; \begin{array}{l} v_{i-1} \text{ is connected by} \leq 1 \text{ edge to the} \\ \text{shortest path from } v_{i-2} \text{ to } v_i \end{array} \right\}$$

*See Figure 4 for a visualization of a MAX-arrangement.*

The key property of a MAX-arrangement is stated by the following two lemmas: consider a node $v_i$ in a MAX-arrangement, then (1) its MAX-arrangement successor node $v_{i+1}$ cannot have a much lower private cost than $v_i$ and (2) in the connection graph the shortest path from $v_i$ to $v_{i+1}$ can overlap by at most one edge with the shortest path to $v_i$'s MAX-arrangement predecessor node.

**Lemma 2.5** *For each two successive nodes $v_i, v_{i+1}$ ($0 \leq i < m$) in a MAX-arrangement $v_0, \ldots, v_m$ it holds $d(v_i, v_{i+1}) \geq c(v_i) - 1$ and hence $c(v_{i+1}) \geq c(v_i) - 1$.*

**Proof.** Consider a node $v_i$, $0 \leq i < m$, in the MAX-arrangement. Then by Lemma 2.3 there exist $x, y \in I(v_i)$ with $d(v_i, x) = c(v_i)$ and $c(v_i) \geq d(v_i, y) \geq c(v_i) - 1$ such that $v_i$ is connected by at most one edge to the shortest path from $x$ to $y$. At least one of these nodes is a valid candidate for the next MAX-arrangement node $v_{i+1}$ (even if neither $x$ or $y$ is $v_{i+1}$, we get a distance lower bound) and we get $d(v_i, v_{i+1}) \geq \min\{d(v_i, x), d(v_i, y)\} \geq c(v_i) - 1$. This gives, $c(v_{i+1}) \geq c(v_i) - 1$. □

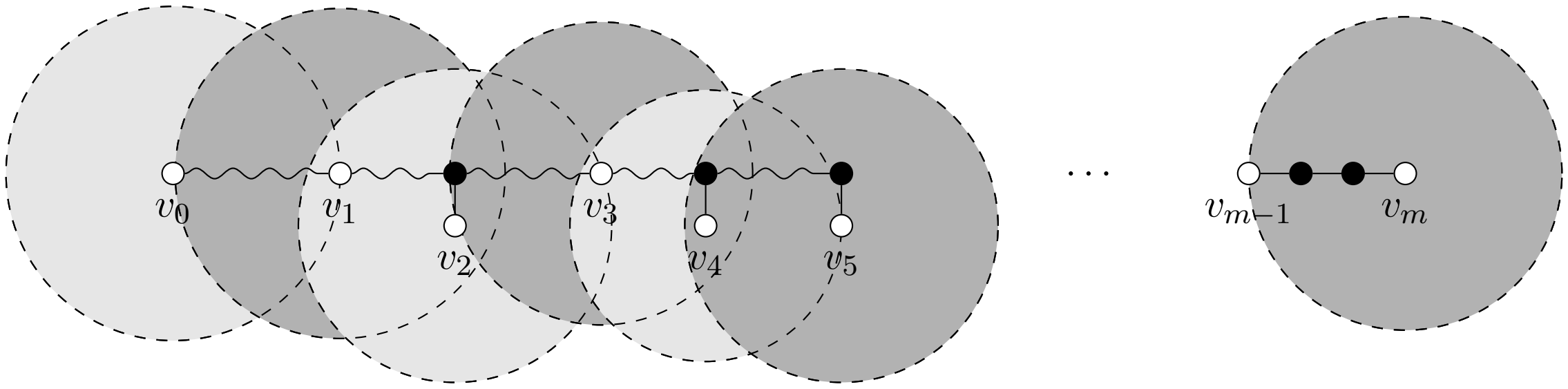

Figure 4: Visualization of a MAX-arrangement. The radius of a circle around a node corresponds to the node's private cost. Curled lines denote shortest paths.

**Lemma 2.6 (Increasing Distance)** *Let $I$ be a set of interests and $G = (V, E)$ a corresponding tree in a MAX-equilibrium with $v_0, \ldots, v_k$ a MAX-arrangement. Then the distances to $v_0$ are monotonously increasing, i.e., $d(v_0, v_i) \leq d(v_0, v_{i+1})$ for $i = 1, \ldots, k-1$.*

**Proof.** By $c(v_1) \geq 3$ we get with Remark 2.2 that $|\, I(v_1)| \geq 2$. Hence by Lemma 2.3 there exists a node $v_2$, such that the paths $v_1$ to $v_0$ and $v_1$ to $v_2$ overlap by at most one edge. By construction of the MAX-arrangement the distance $d(v_0, v_2)$ is maximal among all distances from $v_0$ to nodes $v \in I(v_1)$ and hence we get $d(v_0, v_1) \leq d(v_0, v_2)$.

Assume that there is a node $v_i$ with smallest index $i \geq 2$ in the MAX-arrangement for which the claim does not hold. This is $d(v_0, v_{i-1}) \leq d(v_0, v_i) > d(v_0, v_{i+1})$. Denote by $x$ the most distant node from $v_0$ that is on all shortest paths from $v_0$ to $v_{i-1}$, $v_0$ to $v_i$, and $v_0$ to $v_{i+1}$. (Such a node $x$ exists since especially $v_0$ fulfills the restrictions.) By the choice of $i$ and since all these paths cross node $x$, we get:

$$d(x, v_{i-1}) \leq d(x, v_i) > d(x, v_{i+1}) \tag{1}$$

By definition of the MAX-arrangement, $v_i$ is connected by at most one edge to the shortest path from $v_{i-1}$ to $v_{i+1}$. Hence, $x$ must be a node on the path from $v_{i-1}$ to $v_{i+1}$. First note that $x$ cannot be $v_i$ or a neighbor of $v_i$, since for those cases with (1) we get $d(x, v_{i+1}) < d(x, v_i) \leq 1$. Further, $x$ must lie on the shortest path from $v_{i-1}$ to $v_i$, since otherwise $x$ would lie on the shortest path from $v_i$ to $v_{i+1}$ which implies by $d(v_{i-1}, v_i) \geq 3$ that $d(x, v_i) < d(x, v_{i-1})$. This gives $d(x, v_i) \leq d(x, v_{i+1})$ and is a contradiction. $\square$

**Lemma 2.7** *Let $I$ be a set of interests and $G = (V, E)$ a corresponding tree in a MAX-equilibrium. Consider a MAX-arrangement $v_0, \ldots, v_m$. Then, no edge in $E$ is used more than two times by the shortest path visiting the nodes $v_0, \ldots, v_m$ in the given order.*

**Proof.** We label the nodes of $G$ by their distances to $v_0$. This is, for every $v \in V$ we define a *level* by $\mathrm{level}(v) := d(v_0, v)$. We consider an arbitrary node $v_k$ with $k \in \{1, \ldots, m-1\}$ and the corresponding shortest path $v_k =: w_0 \to w_1 \to \ldots \to w_t := v_{k+1}$ to node $v_{k+1}$. By definition, $v_k$ is connected by at most one edge to the shortest path from $v_{k-1}$ to $v_{k+1}$ (see Figure 3). By Lemma 2.6 we have $\mathrm{level}(v_{k-1}) \leq \mathrm{level}(v_k) \leq \mathrm{level}(v_{k+1})$. Hence, for $i = 2, \ldots, t-1$ we get $\mathrm{level}(w_i) < \mathrm{level}(w_{i+1})$. This is, at most one edge (explicitly edge $\{w_0, w_1\}$) of the shortest path $v_0$ to $v_k$ is used a second time by the shortest path traversal from $v_k$ to $v_{k+1}$. By Lemma 2.5 we have $t \geq c(v_k) - 1 \geq 3$ and get $\mathrm{level}(v_k) < \mathrm{level}(v_{k+1})$. $\square$

Now we prove that given a node $v_0$ with private cost $c(v_0) > 3$ of a MAX-equilibrium tree, there exists a MAX-arrangement that starts with this node and ends with a node with private cost of $3$. With the previous results about MAX-arrangements, this finally leads to the upper bound.

**Lemma 2.8** *Let $I$ be a set of interests and $G = (V, E)$ a corresponding tree in a MAX-equilibrium. Then for $v_0, v_1 \in V$ with $d(v_0, v_1) > 3$ and $v_1 \in I(v_0)$ there exists a MAX-arrangement starting with $v_0$. And for every such MAX-arrangement it holds that the shortest path that visits all nodes of the MAX-arrangement in the given order uses at least $(c(v_0)^2 + c(v_0) - 6)/4$ different edges of $G$.*

**Proof.** *Existence:* $v_0, v_1$ obviously fulfill the conditions for a MAX-arrangement. Thus, it suffices to show that given the beginning of a MAX-arrangement $v_0, \ldots, v_i$ with $c(v_j) > 3, j = 0, \ldots, i-1$ we can either find a next node $v_{i+1}$ that suffices the conditions or otherwise $c(v_i) = 3$. Assume $c(v_i) > 3$. Then, by Lemma 2.3 there exist $x, y \in I(v_i)$ with $d(v_i, x) = c(v_i)$ and $c(v_i) \geq d(v_i, y) \geq c(v_i) - 1$ such that $v_i$ is connected by at most one edge to the shortest path from $x$ to $y$. Since $c(v_i) > 3$, also $c(x) \geq 3$ and $c(y) \geq 3$ hold. Now, for at least one node ($x$ or $y$) we have that this node is most distant to $v_{i-1}$, it is not $v_{i-2}$, and thus it fulfills the conditions for a MAX-arrangement.

*Traversal:* Given the existence, we now can apply the previous lemmas for providing the minimal length of such a MAX-arrangement: Lemma 2.5 states that by construction of the MAX-arrangement we always have $c(v_{i+1}) \geq c(v_i) - 1$. Lemma 2.6 implies that no node can be contained more than once in a MAX-arrangement. By the arguments above we get that we always can find a new node for the MAX-arrangement until we reach a node $w$ with $c(w) = 3$. Hence, the MAX-arrangement contains at least $c(v_0) - 2$ nodes. Since the distance between two succeeding nodes in the MAX-arrangement decreases by at most one per node, a traversal of this MAX-arrangement consists of at least $\sum_{i=3}^{c(v_0)} i = (c(v_0)^2 + c(v_0) - 6)/2$ edges. From these edges, by Lemma 2.7, at least $(c(v_0)^2 + c(v_0) - 6)/4$ edges are different. □

**Theorem 2.1 (Restated)** *Let $I$ be a set of interests and $G = (V, E)$ a corresponding tree in a MAX-equilibrium, $n := |V|$. Then, for all $v \in V$ we have $c(v) \in \mathcal{O}(\sqrt{n})$.*

**Proof.** W.l.o.g. we may assume that there is at least one interest $\{v, v'\} \in I$ with $d(v, v') \geq 3$. Let nodes $v_0, v_1 \in V$, $v_1 \in I(v_0)$ have maximal distance among all nodes, $D := d(v_0, v_1) = c(v_0)$. Then, by Lemma 2.8 we can find a MAX-arrangement $v_0, \ldots, v_m$ whose traversal uses at least $(D^2 + D - 6)/4$ different edges. Since our tree has exactly $n - 1$ edges, we get $(D^2 + D - 6)/4 \leq n - 1$ as an upper bound for the size of every MAX-arrangement and hence the private cost upper bound is $D \in \mathcal{O}(\sqrt{n})$. □

## 2.2 The Private Cost Upper Bound is Tight

The proven upper bound of $\mathcal{O}(\sqrt{|V|})$ for the private costs is also tight, i.e., there exist interest graphs and corresponding tree networks in a MAX-equilibrium with at least one player having private cost $\Omega(\sqrt{|V|})$. For this, we consider an interest graph that forms a circle.

**Remark 2.9** *Let $I := \{\{v_i, v_{i+1}\} | i = 1, \ldots, n-1\} \cup \{\{v_n, v_1\}\}$ be a set of interests for nodes $V := \{v_1, \ldots, v_n\}$, such that $(V, I)$ forms a circle. Consider a connection graph $G$ in which a node $v_i$ has degree one. Then $v_i$ cannot perform any swap if and only if it holds $|d(v_{i-1}, v_i) - d(v_i, v_{i+1})| \leq 1$ and $v_i$ is connected by one edge to the shortest path from $v_{i-1}$ to $v_{i+1}$ (hence $v_i$ is in a T-configuration, cf. Lemma 2.3).*

**Theorem 2.10** *There exists a set of interests and a corresponding tree $G = (V, E)$ in a MAX-equilibrium with a node $v_i \in V$ that has private cost $c(v_i) \in \Omega\left(\sqrt{|V|}\right)$.*

**Proof.** For nodes $v_1, \ldots, v_n$ consider the interests $I := \{\{v_i, v_{i+1}\} | i = 1, \ldots, n-1\} \cup \{\{v_n, v_1\}\}$ and a corresponding connection graph $G = (V, E)$ as stated in Figure 5. We claim that this graph is in a MAX-equilibrium yielding private cost of $c(v_i) = \mathcal{O}(\sqrt{n})$ for a node $v_i \in V$. The graph is designed such that the

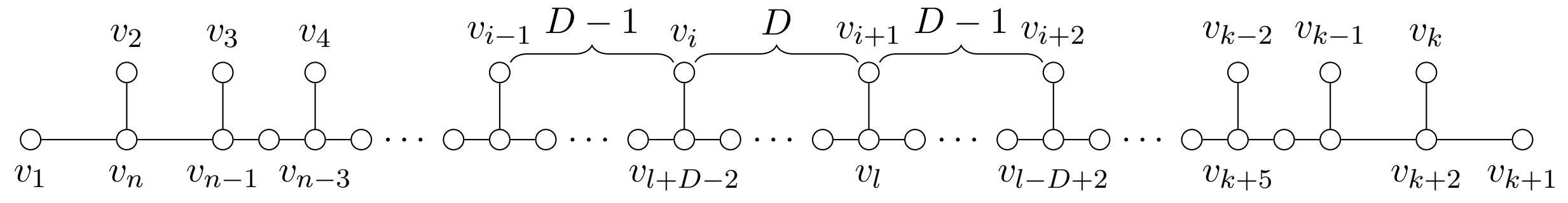

Figure 5: Tree $G = (V, E)$ in a MAX-equilibrium with private cost $\Omega(D)$ for node $v_i$, with $D := \sqrt{|V| - 2} + 1$, $k := 2D - 3$, and $l = n - \sum_{i=1}^{D} i$.

private costs of nodes $v_1, \ldots, v_i$ increase by one each (i.e., for $j = 1, \ldots, i-1$: $c(v_{j+1}) = c(v_j) + 1$) and from $v_{i+1}$ to $v_{k+1}$ decrease by one each (i.e., for $j = i+1, \ldots, k$: $c(v_j) = c(v_{j+1}) + 1$). Node $v_i$ has the maximal private cost. We first compute the exact value for $c(v_i)$ given by this setting, then we argue why no node in this graph can perform an improving step.

Denote the maximal distance from $v_i$ to an interest of $v_i$ by $D$. Then $D$ must fulfill $n = \sum_{i=1}^{D-2} i + \sum_{i=1}^{D-3} i + 2(D-2) + 3 = D^2 - 2D + 3$. Hence, $D = \sqrt{n-2} + 1$. This yields private cost of $\sqrt{n-2} + 1$ for node $v_i$ and gives for the parameters $i := D - 1$ and $k := 2D - 3$.

For each node with degree greater than 1 in $G$ we have private cost of 1 and hence no improving step is possible. For each node with degree equals 1 in $G$ we get by Remark 2.9 that those nodes also cannot perform an improving step. □

## 2.3 Existence of MAX-equilibria and the Price of Anarchy

In this section we consider the social cost of a MAX-equilibrium, this is, we compute the *price of stability* (PoS) and the *price of anarchy* (PoA). Let the *social optimum* represent the smallest social cost of any tree over all nodes (which is not necessarily in a MAX-equilibrium). Then, the *price of stability* denotes the ratio between the minimum social cost of a MAX-equilibrium and the cost of a social optimum. Whereas the *price of anarchy* denotes the ratio between the worst social cost of a MAX-equilibrium and the cost of a social optimum.

We show that there always exists a MAX-equilibrium whose social cost is at most twice as bad as a social optimum. Furthermore, we derive an upper bound for the social cost from the private cost upper bound and show that this bound is tight. Eventually, we can conclude that the price of anarchy is $\Theta(\sqrt{n})$ and the price of stability is at most 2.

**Lemma 2.11** *For every set of interests $I$ there exists a corresponding tree $G = (V, E)$ in a MAX-equilibrium that causes social cost $c(G) \le 2n$, $n := |V|$.*

**Proof.** For the given set of interests we construct a graph by application of Algorithm 1. We argue that the computed graph is a tree and that this tree is in a MAX-equilibrium.

In the first loop (lines 4–6) we only add edges between two nodes $v$ and $w$ that are interested in each other and where one of these nodes has exactly one interest. Thus, the resulting graph at line 6 is acyclic. Note that for every node $v \in B$ that is considered in the second loop (lines 7–12) it holds: (1) $v$ is interested in at most one node to that it is not yet connected and (2) $v$ is connected to a node $u \in A$ (otherwise $v$ would have degree one in the interest graph and therefore would be an element of $A$ instead of $B$). Thus, adding an edge $\{v, w\}$ with $v \in B$ and $I(v) \backslash \{u \in I(v) \mid \exists u' \ (u, u') \in E\} = \{w\}$ in lines 9 does not create cycles in the graph. Finally, in the next loop (lines 14–16) we create a star that connects all formerly created disjoint trees of the graph and by this, the constructed graph $(V, E)$ is a set of trees. In the final loop (lines 17–20), we connect all remaining unconnected components.

It remains to show that $(V, E)$ is in a MAX-equilibrium. By construction, each node added during the first two loops (lines 4–12) has private cost of 1. The center node $x$ (selected in line 13) of the star also has

**Algorithm 1:** Computation of MAX-equilibrium with social cost of at most $2n$

1 $E \leftarrow \emptyset$
2 $A \leftarrow \{v \in V \mid |I(v)| = 1\}$
3 $B \leftarrow V \setminus A$
4 **foreach** $v \in A$ **do**
5 $\quad E \leftarrow E \cup \{\{v, w\} \mid I(v) = \{w\}\}$ $\qquad \triangleright$ note that $w$ exists and is unique
6 **end**
7 **while** $\exists\, v \in B$ *with* $\#\{w \in I(v) \mid \{v, w\} \notin E\} \leq 1$ **do**
8 $\quad$ **if** $\#\{w \in I(v) \mid \{v, w\} \notin E\} = 1$ **then**
9 $\qquad E \leftarrow E \cup \{\{v, w\} \mid w \in I(v),\ \{v, w\} \notin E\}$ $\qquad \triangleright$ note that $w$ exists and is unique
10 $\quad$ **end**
11 $\quad B \leftarrow B \setminus \{v\}$
12 **end**
13 select an arbitrary node $x \in B$ and assign $B \leftarrow B \setminus \{x\}$
14 **foreach** $w \in B$ **do**
15 $\quad E \leftarrow E \cup \{\{x, w\}\}$
16 **end**
17 **foreach** *connected component* $C \subset (V, E)$ *with* $x \notin C$ **do**
18 $\quad$ select an arbitrary node $y \in C$
19 $\quad E \leftarrow E \cup \{x, y\}$
20 **end**
21 **return** $E$;

private cost of 1, since each interest of $x$ is either connected to $x$ in the previous two loops (lines 4–12) or it is connected to $x$ in the last loop (lines 14–16). For each interest $u$ of a node $w \in B$ that is chosen in line 14 it holds: Either $u$ is chosen in the previous loops (lines 4–12) and the edge $\{u, w\}$ is added to $E$, or $u$ is connected to $x$ in the last loop (lines 14–16). In both cases, the distance from $w$ to $u$ is at most 2. Thus, $w$ has private cost of 2. Since $w \notin A$, we have $|I| > 1$ and thus, $w$ cannot perform an improving step. In total, we get private cost of at most 2 for each node. □

**Theorem 2.12** *The price of stability for I-BNCG is at most* 2*.*

**Proof.** Let $I$ be a set of interests over nodes $V$, $n := |V|$. Then each connection graph that is a tree induces social cost of at least $n$. By Lemma 2.11 there exists a connection graph in a MAX-equilibrium with social cost of at most $2n$. Thus, the price of stability is at most $2n/n = 2$. □

Next, we prepare our estimation of the price of anarchy.

**Lemma 2.13** *There exist interest graphs over $n$ nodes with a corresponding MAX-equilibrium tree that causes social cost of* $\Omega\left(n \cdot \sqrt{n}\right)$.

**Proof.** For nodes $V = \{v_1 \ldots, v_n\}$ consider the interests

$$\begin{aligned} I :=& \{\{v_i, v_{i+1}\} | i = 1, \ldots n/2 - 1\} \cup \{\{v_i, v_1\} | i = n/2, \ldots, n\} \cup \\ & \{\{v_{n/2-1}, v_i\} | i = n/2, \ldots, n\} \end{aligned}$$

as stated in Figure 6. Further, we consider the connection graph $G = (V, E)$ (stated in Figure 7) where the nodes $n/2, \ldots, n$ are all connected to node $v_l$. For the distance $C$ between node $v_l$ and $v_{l+C}$ it must hold: $n = 2\sum_{i=1}^{C} i + \frac{n}{2} + 2C + 4$. Hence, we have $C = \frac{\sqrt{2n-7}-3}{2}$. This implies private cost of $\frac{\sqrt{2n-7}+1}{2}$ for all nodes $v_{\frac{n}{2}}, \ldots, v_n$. Thus, the social cost of $G$ is $\Omega(n \cdot \sqrt{n})$.

It remains to show that $G$ is in a MAX-equilibrium. But this argument is analog to the proof of Lemma 2.10. □

**Theorem 2.14** *The price of anarchy for I-BNCG is $\Theta(\sqrt{n})$, with $n$ being number of nodes in the graph.*

**Proof.** Theorem 2.1 provides an upper bound of $\mathcal{O}\left(\sqrt{n}\right)$ for the private cost of every node in a tree in a MAX-equilibrium with $n$ nodes. By this we get $\mathcal{O}\left(n \cdot \sqrt{n}\right)$ as an upper bound for the social cost of every MAX-equilibrium. Further, by Lemma 2.13 we get $\Omega\left(n \cdot \sqrt{n}\right)$ as a worst case lower bound for the social cost of a graph in a MAX-equilibrium. For the social optimum we can get $\Theta\left(n\right)$. (Social optimum incurs cost of at least $n$ and at most $2n$.) Hence, by combination of these results we get $\Theta\left(n \cdot \sqrt{n}/n\right) = \Theta\left(\sqrt{n}\right)$ for the price of anarchy. □

### 2.4 The Price of Anarchy for I-BNCG on General Graphs

In contrast to I-BNCG on trees, the price of anarchy for general connection graphs is $\Theta\left(n\right)$.

**Theorem 2.15** *The price of anarchy for I-BNCG with arbitrary connection graphs is $\Theta\left(n\right)$, with $n$ being number of nodes in the graph.*

**Proof.** First, note that the social costs of every instance are upper bounded by $n^2$ and lower bounded by $n$. Second, we provide an interest graph over $n$ nodes ($n \equiv 0 \mod 6$) and a corresponding MAX-equilibrium graph $G = (V, E)$ with social cost $\Omega\left(n^2\right)$ (see Figure 8). To this, we connect $n/2$ nodes to a ring (ring nodes) and connect one additional (satellite) node to each of them. Each of the ring nodes is interested in its three adjacent nodes in $G$, whereas each satellite node is interested in its neighbor at the ring and in both satellite nodes at distance exactly $n/6 + 2$. This is an equilibrium and all $n/2$ satellite nodes have private cost $n/6 + 2$. This gives a price of anarchy of $\Omega\left(n\right)$. □

## 3 Further Structural Properties of Equilibria

In this section we present for the MAX-version how the size of a maximum independent set in the interest graph limits the maximal private cost of every node and by this the maximal social cost of the graph. Furthermore, we show that in the AVG-version there are equilibria with nodes having private cost linear in the number of nodes.

### 3.1 Cost Limitation for MAX-Version by MIS in the Interest Graph

By Lemma 2.8 we achieved a general upper bound for any MAX-arrangement (see Definition 2.4) contributed only by the property that the network is connected. Here, we introduce a second upper bound for a MAX-arrangement that is given by the size of a *maximum independent set* (MIS) in the interest graph. Having such a MIS of size $M$, we can bound the maximum private costs by $\mathcal{O}\big(M\big)$ which yields improved bounds for specific families of interest graphs.

**Theorem 3.1** *Let $I$ be a set of interests and $G = (V, E)$ a corresponding tree in a MAX-equilibrium. Let $M$ be the size of a maximum independent set in $(V, I)$. Then for every MAX-arrangement $v_0, \ldots, v_m$ we get: The length of this MAX-arrangement is at most $2 \cdot M$.*

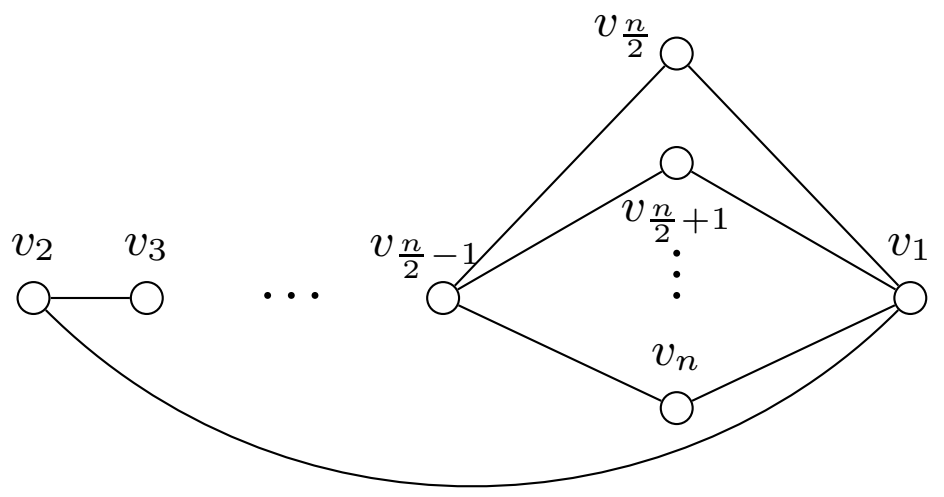

Figure 6: The interest graph for the proof of Lemma 2.13.

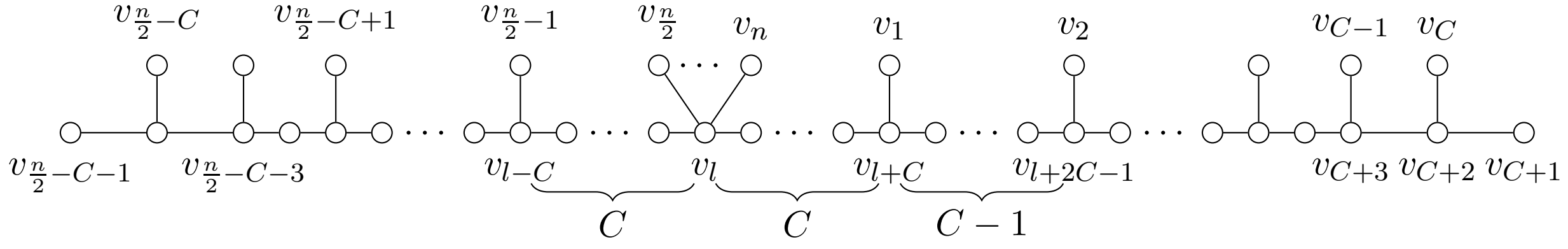

Figure 7: The connection graph for the proof of Lemma 2.13. This tree corresponds to the interests as given in Figure 6. The parameters are $C = \frac{\sqrt{2n-7}-3}{2}$ and $l = \frac{n}{2} - C - (\sum_{i=1}^{C} i + 2)$.

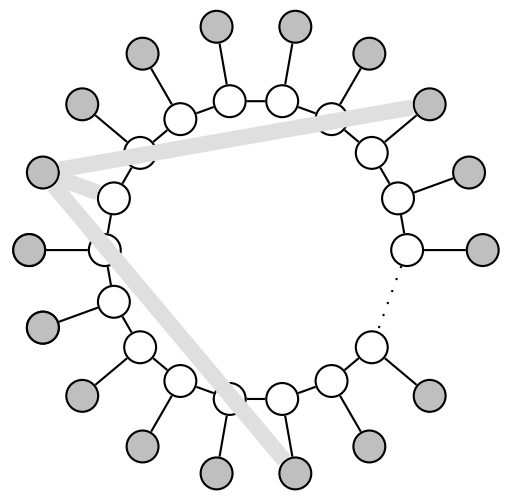

Figure 8: Lower bound construction for the price of anarchy for general graphs. Each of the white nodes is interested in its three neighbors. Each gray node is interested in its white neighbor and the two gray nodes at distance exactly $n/6 + 2$.

**Proof.** We prove that the nodes of $v_0, \dots, v_{m-1}$ with even index form an independent set in the interest graph $(V, I)$. Consider an even index $i$ and assume for contradiction that there is an even index $k < i$ such that $v_k \in I(v_i)$. By Lemma 2.6 we get $d(v_k, v_{k+1}) \leq d(v_k, v_{k+2})$. If $v_{k+2} \neq v_i$ with Lemma 2.5 and $c(v_j) > 3$ for all $v_j$ in the MAX-arrangement we get $d(v_k, v_i) > d(v_k, v_{k+2}) + 1 \geq c(v_k)$. But this is a contradiction.

Thus, consider the case $v_{k+2} = v_i$. Since $v_{k+1}$ is connected by at most one edge to the shortest path from $v_k$ to $v_{k+2}$ and $d(v_{k+1}, v_{k+2}) \geq 3$ we get that $v_{k+2} \notin I(v_k)$. Otherwise we either get the same contradiction as before, or $v_{k+1}$ would contradict to be the most distant node in $I(v_k)$ that fulfills the MAX-arrangement conditions.

Hence, the nodes with even index of the MAX-arrangement form an independent set in $(V, I)$. Since an independent set has at most $M$ nodes, we get an upper bound of $2 \cdot M$. □

**Corollary 3.2** *Let $I$ be a set of interests and $G = (V, E)$ a corresponding tree in a MAX-equilibrium, $n := |V|$. Let the size $M$ of any MIS in $(V, I)$ be limited by $\sqrt{n}$. Then, for $v \in V$ we have $c(v) \in \mathcal{O}(M)$.*

**Proof.** W.l.o.g we assume that there is a node with private cost greater than 3. Hence, there exists a MAX-arrangement $v_0, \dots, v_m, v_{m+1}$ with $c(v_i) > 3, i = 1, \dots, m$ and $c(v_{m+1}) = 3$. By Theorem 3.1 we get $m \leq 2M$. Analog to Theorem 2.1 we get the upper bound. □

Furthermore, our technique directly yields asymptotically same private cost upper bounds for complete interest graphs on trees as those that were explicitly constructed by Alon et al. [2].

**Corollary 3.3** *Let $I$ be a set of interests and $G = (V, E)$ a corresponding tree in a MAX-equilibrium, $n := |V|$. If $(V, I)$ is a complete graph, then the maximal private cost of every node is $\mathcal{O}(1)$.*

We conclude our analysis of MAX-equilibria with a negative result. This is, even if there always exists a MAX-equilibrium there are possible invocation sequences of the nodes such that the connection graph never reaches an equilibrium.

**Remark 3.4** *There are invocation sequences that contain all nodes of a connection graph, where on invocation each node performs a best-response improving swap (if possible), such that the nodes never reach a MAX-equilibrium. An explicit construction of such an instance is given by Figure 9.*

From Remark 3.4 we can conclude that there does not exist a potential function for general I-BNCGs. Thus a I-BNCG is not a potential game as defined by Monderer and Shapley [11].

### 3.2 Linear Private Cost Lower-Bound for AVG-Version

Here, we consider the AVG-version and provide an equilibrium where a node's private cost can be lower bounded by the number of nodes in the connection graph. In the AVG-version, the private cost of a node is defined by $c(v) := \sum_{w \in I(v)} d(v, w)/|\,I(v)|$. In comparison, for complete interest graphs Alon et al. proved [2] that every AVG-equilibrium tree has a diameter of at most 2, hence each node's private cost is bounded by 2.

**Theorem 3.5** *There exists a set of interests $I$ and a corresponding tree $G = (V, E)$ in an AVG-equilibrium, such that $G$ contains a node $v_i \in V$ that has private cost $c(v_i) = \Omega(|V|)$.*

**Proof.** We consider the graph as stated by Figure 10 with $n := |V|$. The graph is given by nodes $V = \{v_1, \dots, v_n\}$ and edges $E := \{\{v_i, v_{i+1}\} | i = 1, \dots, n-1\}$. We consider the set of interests $I := \{\{v_i, v_{i+1}\} | i = 1, \dots, n-1\} \cup \{\{v_n, v_1\}\}$. All nodes $v_2, \dots, v_{n-1}$ have private cost of 1, which is minimal and gives that these nodes cannot perform an improving step. For $v_1$ we get that swapping its edge to any other node would increase the distance to $v_2$ by the same value as the distance to $v_n$ would decrease. (For $v_n$ this is the symmetric case.) Hence, neither $v_1$ nor $v_n$ can perform an improving step and the tree is in AVG-equilibrium with $c(v_1) = n/2 = \Omega(n)$. □

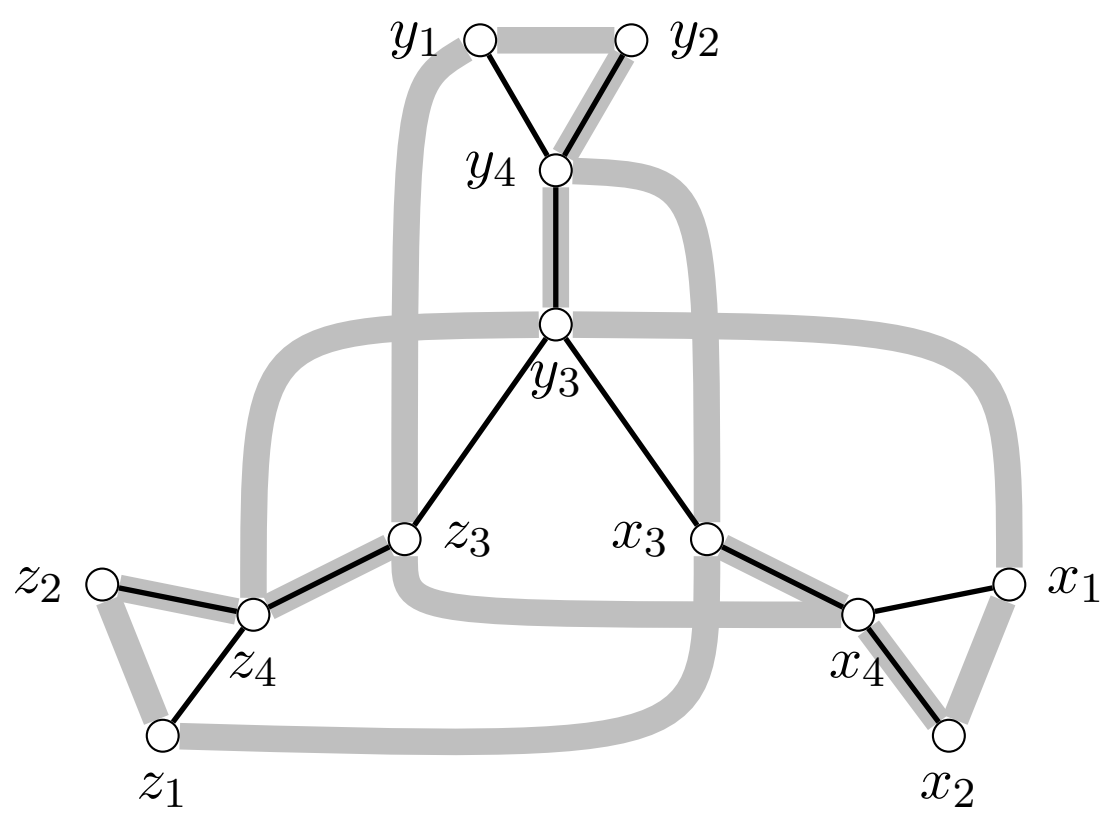

Figure 9: The gray lines are the interests, the black lines the connection graph. Consider the invocation sequence $x_1, \ldots, x_4, y_1, \ldots, y_4, z_1, \ldots, z_4$. On invocation, every node performs a best response swap in the MAX-version. By this, only the following swaps are performed: $x_3 : [y_3 \to z_3], y_3 : [z_3 \to x_3], z_3 : [x_3 \to y_3]$ (in this order). After one sequence the connection graph is again the initial connection graph. By repeating this sequence a MAX-equilibrium is never reached.

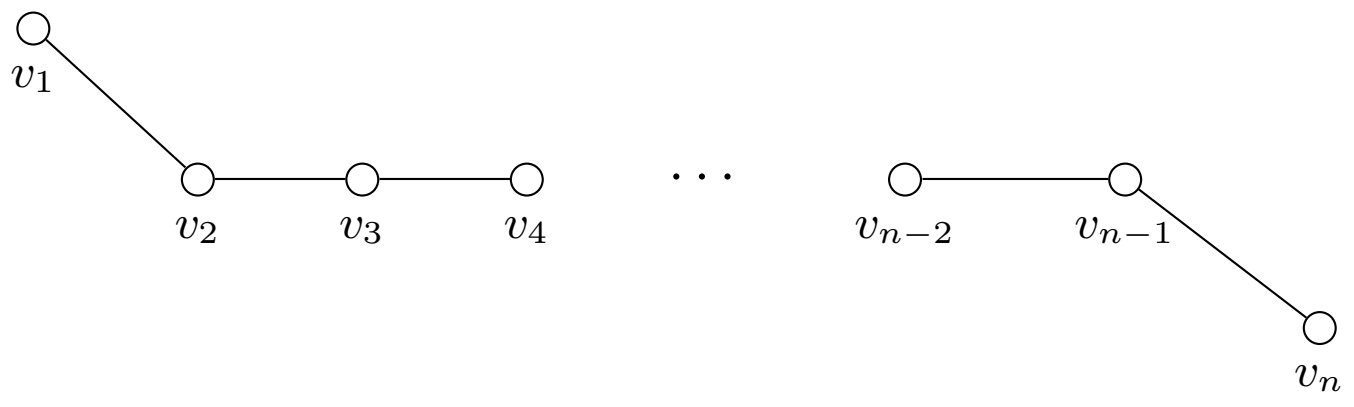

Figure 10: Tree $G = (V, E)$ in AVG-equilibrium with private costs $\Omega(|V|)$ for nodes $v_1$ and $v_n$.

# 4 Outlook and Future Work

In this paper, we presented tight worst case bounds for the private costs as well as for the social cost in any MAX-equilibrium on tree networks. Furthermore, we drew an interesting connection between the size of an MIS in the interest graph and upper bounds on the private/social costs. In comparison with MAX-equilibria on general graphs, we could show that the price of anarchy can perform much worse if the connection graph is not acyclic. However, it remains an open question whether the price of anarchy on general connection graphs with complete interests could perform better than $\mathcal{O}(n)$. For this, so far there is only a worst case lower bound of $\Omega(\sqrt{n})$ (by Alon et al. [2]) for the graph diameter in a MAX-equilibrium, yielding a lower bound for the price of anarchy. Techniques similar to our MAX-arrangement-technique may allow deeper insights into the nature of MAX-equilibria in that scenario. Apart from this, finding good upper bounds on the social cost of an AVG-equilibrium remains a challenging problem (we gave a lower bound of $\Omega(n)$ for the private costs).

Even if the existence of a MAX-equilibrium is always ensured (which we proved for the MAX-version on trees), it remains an open question whether the process ever reaches an equilibrium. We could state for the MAX-version examples of interest graphs and corresponding invocation sequences of the nodes, where the network *never* converges to a MAX-equilibrium. It seems an interesting question whether we can guarantee the convergence by additional policies, e.g., by restricting the order in which nodes perform their swaps. And in the case of a guaranteed convergence, how many swaps would it take to reach an equilibrium?

Currently, we only considered static interest graphs: the set of interest never changes. In practice, interests of network participants might change over time. Introducing a discrete time model and considering certain (possibly restricted) changes of the interest graph seems a natural way to generalize our model, yielding an interesting online problem.